\documentclass[aps,prb,twocolumn,superscriptaddress,floatfix]{revtex4}

\usepackage{amsmath}
\usepackage{subfigure}
\usepackage{xcolor}
\usepackage{graphicx}

\newcommand\tikzinput[1]{ \includegraphics{#1.pdf}}

\newcommand{\bfr}{\textbf{r}}
\newcommand{\bfR}{\textbf{R}}

\newcommand{\Pri}{P_{\langle  i\rangle}}
\newcommand{\Prj}{P_{\langle  j\rangle}}

\newcommand{\cj}{c_{j}}
\newcommand{\cdagi}{c^{\dagger}_{i}}

\newcommand{\bfu}{{\textbf u}}
\newcommand{\bfv}{{\textbf v}}
\newcommand{\bfK}{{\textbf K}}
\newcommand{\etal}{\emph{et al.}}

\newcommand{\lead}{{\textrm{lead}}}
\newcommand{\maxx}{{\textrm{max}}}
\newcommand{\minn}{{\textrm{min}}}
\newcommand{\subb}{{{\textrm{sub}}}}
\newcommand{\SW}{{\textrm{SW}}}

\begin{document}

\title{ Order and supersymmetry at high filling zero-energy states on the triangular lattice}
\author{D. Galanakis}
\affiliation{School of Physical and Mathematical Sciences, Nanyang Technological University, Singapore 637371}
\author{C. L. Henley}
\affiliation{Dept. of Physics, Cornell University, Ithaca NY 14853-2501 USA}
\author{S. Papanikolaou}
\affiliation{Department of Mechanical Engineering and Materials Science, Yale University, New Haven, Connecticut, 06520-8286, USA}
\affiliation{ Department of Physics, Yale University, New Haven, Connecticut, 06520-8120, USA}
\date{\today}
\begin{abstract}
We perform exact diagonalization studies in $d=2$ dimensions 
for the Fendley and Schoutens model of hard-core and nearest-neighbor 
excluding fermions that displays
an exact non-relativistic supersymmetry.
Using clusters of all possible shapes up to 46 sites,
we systematically study the behavior of the ground state phase diagram 
as a function of filling. We focus on the highly degenerate zero-energy
states found at fillings between $1/7$ and $\sim 1/5$. 
At the lower end of that interval, at filling $1/7$, we explicitly show that the ground states are gapped crystals.  
Consistent with previous suggestions, we find that the extensive entropy of
zero states peaks at a filling of $\sim 0.178$. 
At the higher end of the interval, we find zero energy ground states 
at fillings above $1/5$, in contradiction to previous numerical studies and analytical suggestions; these display non-trivial amplitude degeneracies.
\end{abstract}

\maketitle

\section{Introduction}

Supersymmetry is an extended symmetry which partners together fermions and bosons and is typically studied in the context of extensions of the high-energy Standard model of particle physics~\cite{weinberg-book}. Recently, Fendley and Schoutens~\cite{Fendley:2003uq,Fendley:2003fk} proposed a  many-body lattice model which exhibits what is known as supersymmetric quantum mechanics~\cite{Ilieva:2006qf,Hagendorf:2011bh}. 
It consists of hard-core lattice fermions with nearest neighbor exclusion and a precisely fine-tuned interaction coupling. 
The eigenstates of the model consist of positive doubly degenerate energy states where the supersymmetry is spontaneously ``broken" and  unpaired zero-energy ground states. The nature of these states
was studied extensively on the square lattice~\cite{Huijse:2008zr}, 
in specially decorated two-dimensional lattices\cite{Fendley:2005ys,Huijse:2012},
and also in one-dimensional models, where it provided insights into hidden many-body symmetries~\cite{Fendley:2011ly}.

On the square lattice the number of zero-energy states  was exponential
only in the system's \emph{linear} dimension~\cite{Huijse:2008zr}, leading to a subextensive entropy of possible ground-states. In contrast, on the triangular 
lattice the number of zero energy states appears exponential in the \emph{area}, as recent exact diagonalization (ED) studies have indicated~\cite{Huijse:2011vn}.
Jonsson, by studying finite triangular clusters, 
recently conjectured~\cite{Jonsson:2010kx}, 
that on the triangular lattice, zero-energy states only appear in the interval $1/7\leq f \leq 1/5$, 
where $f$ is the filling (number of fermions/site). 
\footnote{
Such results are based on cohomology 
theory~(Refs.~\onlinecite{Fendley:2003fk} and \onlinecite{Fendley:2005ys}).
The study in Ref.~\onlinecite{Jonsson:2010kx}
focused only on a certain kind of homology cycles, 
the so-called ``cross-cycles", where each such cycle coincides 
with the fundamental cycle of the boundary complex of a cross-polytope. 
In this case, the homology of the independence complex on the triangular lattice
appears tractable, even though it is not exhaustive.
The count of zero-energy states is better understood in the square lattice case,
where it was even shown that a 1-1 correspondence exists between 
the possible zero-energy states and the tilings of the 
2D hard-squares model at negative fugacity $z=-1$.}
Huijse~\etal~\cite{Huijse:2011vn} recently investigated, using ED,  
the nature and number of zero-energy states on the triangular lattice. Their results appear consistent with Jonsson's 
conjecture and the suggestion of exponential zero energy states' degeneracy, but they 
focused on quasi-one dimensional ladders, of width up to four. 

In this paper, we use exact diagonalization applied to \emph{all}
symmetrically inequivalent periodic clusters of various shapes and sizes, to determine the
phase diagram as a function of filling. By combining data from
many different clusters, we are able to obtain curves of
the energy as a function of filling across the whole phase
diagram, and the entropy of zero states as a function of filling, 
across the interval in which they are found.
Our study emphasizes {\it two}-dimensional clusters, wider than 
those of Ref.~\cite{Huijse:2011vn}. 

The outline of our paper and main findings are as follows.
First, in Section~\ref{sec:model}, we introduce the supersymmetric lattice model 
and the parameters of our exact diagonalization calculations.
In Section~\ref{sec:phase_diagram} we present our results on the different phases 
that appear as the density of fermions increases:
a Fermi liquid phase at low filling $f<1/7$, zero-energy states with extensive entropy at 
intermediate fillings, and high-density states (with tendencies to spatial orders) from
$f\approx 1/5$ up to the maximum filling $f=1/3$. 
In Section~\ref{sec:zero-endpoints}, 
we focus on the states found near either endpoint of the zero-energy ground state interval. 
The minimum filling with a zero energy state is 
exactly $f=1/7$ as conjectured~\cite{Jonsson:2010kx}; however, from that filling
up to $f\approx 0.156$, the only zero-energy states are crystal-like and 
gapped states, and have no extensive entropy.
At fillings slightly higher than $1/5$, 
we find that certain zero-energy states exist,
contradicting the conjecture~\cite{Jonsson:2010kx}. 
These states show a tendency
to anisotropic forms of spatial order, and their wavefunctions exhibit surprising regularities,
in that many inequivalent fermion configurations have the same, maximal
amplitude in the wavefunction (Sec.~\ref{sec:magic}).
Finally, Section~\ref{sec:conclusions} contains a summary of our results and 
speculations on how these results may fit into a complete picture of the phase diagram.

\section{Model and method}
\label{sec:model}

The Hamiltonian of the model is based on the definition of the operators $Q=\sum c_{i}^\dagger P_{\langle  i\rangle}$ and $Q^\dagger=\sum c_{i} P_{\langle  i\rangle}$, where $P_{\langle  i\rangle}=\prod_\textrm{j next to i}(1-c_{j}^\dagger c_j)$ form projectors that exclude nearest neighbor occupancy.
Following the basic recipe on constructing the supersymmetric quantum mechanics~\cite{Witten:1982fk}, the Hamiltonian is just
\begin{equation}
H=\{Q^\dagger,Q\}=t \sum_i\sum_\textrm{j next to i} P_{\langle  i\rangle}\cdagi\cj\Prj + V\sum_i\Pri,
\label{eq:model}
\end{equation}
with $t\equiv V$.
The first term in \eqref{eq:model} is a kinetic term of nearest-neighbor hopping (subject to the
hard-core constraint); the second term is a potential energy, equal to the number of
fermions plus the number of vacant sites \emph{not} forbidden by a neighboring fermion,
which effectively includes pairwise farther-neighbor interactions as well as
multi-fermion interactions.
The Hamiltonian \eqref{eq:model} is supersymmetric only when finely tuned to the 
special point $t=V$.  In view of the massive ground-state degeneracy , any change in the Hamiltonian 
would be a singular perturbation leading to a new phase, if the energy spectrum is gapless; hence, the $t=V$ point
in the extended phase diagram is speculated to be a multicritical point at which various 
phases meet~\cite{Huijse:2008zr}.

The model contains an exact ${\mathcal N}=2$ supersymmetry in that 
all eigenstates with nonzero energy $E$ belong to
degenerate \emph{pairs}, $|a\rangle$ with $F$ fermions and $|b\rangle$ 
with $F+1$ fermions, such that $Q|a\rangle = \sqrt{E}|b\rangle$ and $Q^\dagger|b\rangle=\sqrt{E}|a\rangle$,
whereas $Q|b\rangle = Q^\dagger|a\rangle=0$.
The state with an even number 
is considered ``bosonic'' and the one with an odd number is considered
``fermionic''.  An interesting corollary is that if there is a branch
of fermionic elementary  excitations, there must be a corresponding
bosonic branch and vice versa.

Most importantly, there may be ground states $|G\rangle$ of zero energy,
such that $Q|G\rangle=Q^\dagger|G\rangle=0$. 
A thermodynamically \emph{extensive} zero-energy ground state entropy was reported and analyzed in several lattices, 
including integrable chains or ladders~\cite{Fendley:2005ys,Huijse:2008zr} as well as two-dimensional
lattices, of which the triangle lattice is the simplest.
In each model, the zero-energy states are always limited to fillings in the interval
$f\in (f_-,f_+)$, where  $f$ is the number of fermions per site.
These prior studies mostly indicated that the zero-energy states have a strong tendency 
to form crystalline phases.

We diagonalize the model of Eq.~\eqref{eq:model}
on a triangular lattice using clusters of different size and shapes. Each cluster is characterized by the shortest and the next to shortest edge, $\bfv=(v_1,v_2)$ and $\bfu=(u_1,u_2)$ respectively, where the integer coordinates $u_{1,2}$ and $v_{1,2}$ are expressed relative to the triangular lattice basis vectors $\hat{e}_1=\hat{x}$ and $\hat{e}_2=-\frac{1}{2} \hat{x}+\frac{\sqrt{3}}{2} \hat{y}$; the number of sites
is $N=u_1 v_2- u_2 v_1$.   We used \emph{all} the symmetrically inequivalent clusters 
such that $\textbf{v}$ belongs to the irreducible wedge of the triangular lattice and 
$|\textbf{v}|^2 \leq |\textbf{u}|^2$ for up to $N=46$ sites (larger for some fillings).
We perform the diagonalization separately for each sector defined by fermion count $F$ and 
center of mass momentum. For the diagonalization we use the ARPACK package~\cite{arpack}, 
which is an implementation of the Implicitly Restarted Lanczos Method (IRAM).
Note that in accordance to Ref.~\onlinecite{Huijse:2011vn},
well within the zero state interval we find that every momentum sector 
has a similar number of zero energy states, about 100 per sector
in our largest system ($N=46$); it is this huge degeneracy that limited the cluster size we could handle in this range of fillings. For other fillings,
namely $f<1/7$ and $f>1/5$ we study up to $N=56$ and $N=60$ respectively.
In any case, for comparison, the tractable clusters are typically much larger than in Hubbard model exact diagonalization studies, since there is only one spin species and a hardcore constraint is enforced, 
each of which greatly reduces the Hilbert space.

\section{Results: phase diagram}
\label{sec:phase_diagram}

\begin{figure}
\tikzinput{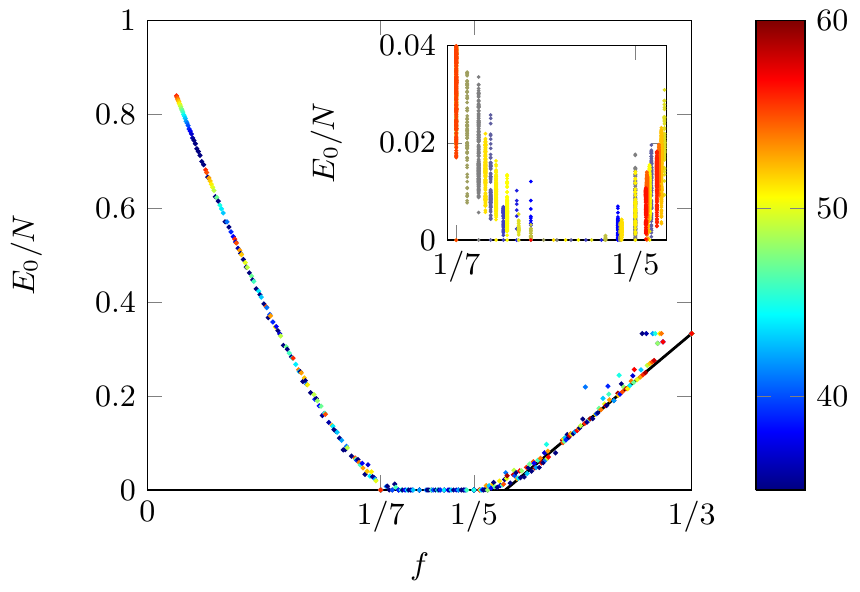}
\caption{ Ground state energy per site vs. filling, minimized over all possible clusters with sizes $35 \leq N \leq N_{\max}(f)$, where $N_{\max}(f)=56$ for $f<1/7$, 46 for $1/7\leq f \leq 1/5$ and 60 for $f>1/5$. To filter out ladders we also impose the constraint $|\bfv| \geq 4$. For each filling the energy is minimized across all clusters and momenta. The different data points are colored according to the total system size. The solid line is a linear low envelop of the high energy regime which has the form $E=\frac{1}{3}+\lambda \left(f-\frac{1}{3}\right)$, with $\lambda=2.92$. In the inset we show the energy versus filling for all the clusters zooming in the interval
of zero energy states.
\label{fig:min_energy}}
\end{figure}

\begin{figure}
\subfigure[$f\rightarrow 0$]{\tikzinput{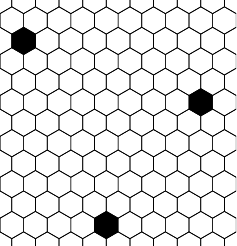}}\subfigure[$f=1/7$]{\tikzinput{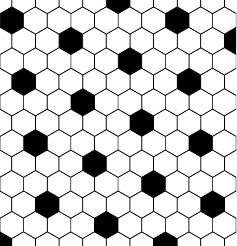}}\subfigure[$f=1/6$]{\tikzinput{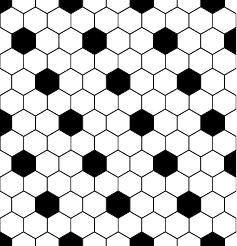}}
\par
\subfigure[$f=1/5$]{\tikzinput{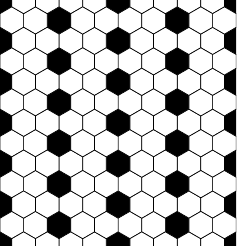}}\subfigure[$f=1/4$]{\tikzinput{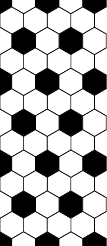}\tikzinput{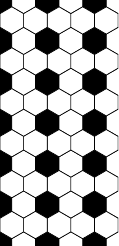}}\subfigure[$f=1/3$]{\tikzinput{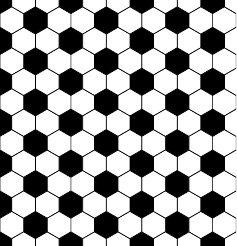}}
\caption {\textbf{Idealized fermion configurations representing phases at various fillings.}
In this and subsequent figures, the triangular lattice sites are the
centers of the hexagons, and filled hexagons are spinless fermions.
For each filling, a configuration of minimal potential energy is shown.
More generally, depending on the unit cell constraints of each 
particular cluster, the configurations of highest weight and
having minimum potential energy tend to be mixtures of triangles
from the crystals shown here.
(a) At low density the system is a dilute gas of weakly interacting fermions. 
(b) A triangular superlattice crystal at $f=1/7$ minimizes the potential 
energy for that filling. 
(c) A covering configuration at $f=1/6$
(d) Two possible covering configurations for $f=1/4$
(e) Anisotropic triangular lattice at $f=1/5$ 
(f) At $f=1/3$, there are only three allowed configurations in which one of 3 sublattices is fully occupied
and no hopping is possible.}
\label{fig:typical}
\end{figure}

The model of Eq.~\eqref{eq:model} displays three clear regimes as a function of particle density 
as evidenced from the equation of state in Fig.~\ref{fig:min_energy} and in the
configurations illustrated in Fig.~\ref{fig:typical}. The qualitative behavior at low and high density follows the studies by Henley and Zhang\cite{Zhang:2003cr,Henley:2001dq}, who previously used exact diagonalization to study a model which was similar in that it has the same Hilbert space -- 
spinless fermions with nearest-neighbor exclusion -- but having $V=0$ in Eq.~\eqref{eq:model}, 
so it lacked supersymmetry. 
At low densities a Fermi liquid phase appears, followed by zero energy ground states for fillings $1/7<f<0.2174$.
At the maximum allowed density of $f=1/3$, the only allowed configuration is a crystalline phase with no quantum fluctuations,
which, dominates the phase diagram beyond the interval of zero-energy states. In the rest of this section
we discuss those three different regimes of the phase diagram.

\subsection{Fermi liquid phase ($f<1/7$)}
\label{sec:fermi-liquid}

The low density Fermi liquid is similar to the one studied by Henley and Zhang\cite{Zhang:2003cr,Henley:2001dq} in their non-supersymmetric model.
The ground state for each filling, at low density,
has the quantum numbers and is approximated by the 
corresponding ground state for non-interacting fermions
on the same cluster.
A consequence of the positive hopping $t>0$ in 
our Hamiltonian is that the single fermion dispersion has \emph{two} 
symmetry-related valleys, with minima at the Brillouin zone corners.
For a given cluster, the sequence of energies as fermions
are added one by one shows ``shell filling effects'' as expected 
in a Fermi liquid~\cite{Fradkin:1991fk}: that is,
in the dilute limit two successive fermions have a similar 
addition cost (placing one in each valley), then the next two are
higher. Thus the addition energies show an even/odd alternation on top 
of an overall increasing trend, due to the repulsive (hardcore) interaction.
If it were necessary to incorporate interaction effects
in a systematic way, one relatively simple approach would be
the $t$-matrix formalism of Ref.~\onlinecite{Zhang:2004}.

\subsection{Zero-energy states ($1/7 < f < f_+$)}
\label{sec:zero+entropy}

\begin{figure}
\tikzinput{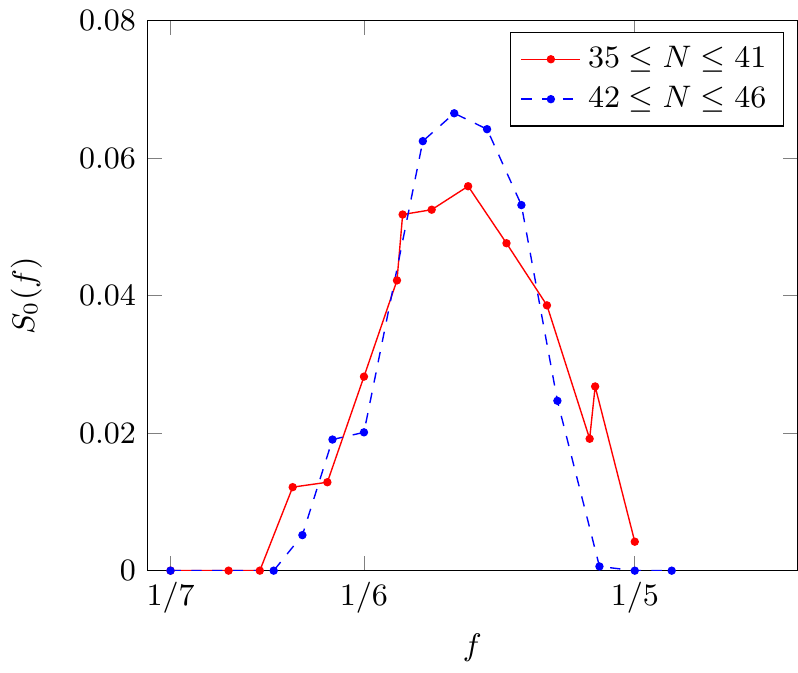}
\caption{
\textbf{Zero energy states entropy}.
The zero state entropy per site, averaged over all the clusters with the same filling 
$f$ for different subsets of clusters.  We imposed the constraint $|\bfv|\geq4$
to filter out the ladders which have a zero energy state at lower fillings. 
The solid line corresponds to all cluster with $N \geq 34$ the solid green 
 to clusters with $35 \leq N\leq 41$ and the dot dashed line to $35 \leq  N\leq 46$. 
 There seems to be a maximum in the vicinity of $f=0.178$.}
 \label{fig:zero-energy-entropy-average}
\end{figure}

The degeneracy of the zero-energy states 
is best characterized by their zero state entropy~\cite{Fendley:2009bh},
which is extensive in the intermediate fillings.
This fact suggests that at the infinite system size limit,
the wave function incorporates a mix of local patterns
(especially fragments of the crystals in Figure~\ref{fig:typical}.
However, at finite size particular periodic boundary conditions may 
be commensurate with only a subset of these;
hence, different clusters may show contrasting behaviors, which
would \emph{all} be present in different patches of a typical infinite-size
configuration.

Let $A_0(F;\mathcal{C})$ be the number of zero energy states 
for each given cluster $\mathcal C$; this is one of the outputs 
of the ED.
We conjecture that $A_0(F;\mathcal{C})\sim \exp(N S_0(f))$,
where $N(\mathcal C)$ is the number of sites in the cluster
and $f=F/N$, with an entropy function 
$S_0(f)$ that is well-defined in the thermodynamic limit.
We construct a numerical approximation to the entropy function
by first computing an entropy per site for each cluster and then 
averaging over all clusters (of any size) with the identical
filling per site, that is 
\begin{equation}
S_0(f)= \left< \frac{1}{N(\mathcal C)}\log\left[N_0\left( f N(\mathcal C);\mathcal C \right) \right] \right>_{\mathcal C}.
\label{eq:entropy}
\end{equation}
This averaging is meant to reduce the commensurability effects 
of different cluster shapes.

To better approximate the entropy of a two dimensional system we excluded 
ladder-shaped clusters by putting a lower limit for the width of $\bfv$. 
We found that ladder-like clusters tend to have 
larger entropies per site than the more two-dimensional ones,
and also this entropy tends to peak at a smaller filling $f$.
Including such clusters would have incorrectly biased our
estimates and can spuriously suggest that $S_0(f)$ has 
another peak below $f=1/6$.

The result is shown in Fig.~\ref{fig:zero-energy-entropy-average},
 where we can can observe four interesting numerical trends
\begin{itemize}
%%%%%%%%%%%%%%%%%%%%%%%%%%%%%%
\item[(1)]
There are no zero-energy states below $1/7$, 
verifying recent expectations~\cite{Jonsson:2010kx} and in agreement with recent numerical exact diagonalization studies.~\cite{Huijse:2011vn}
%%%%%%%%%%%%%%%%%%%%%%%%%%%%%%
\item[(2)] In the interval
$1/7<f<0.156$ almost all clusters either have no zero energy states or have
unique zero energy states in isolated momentum sectors. (In some cases there 
are doubly degenerate zero energy states due to rotational symmetries.)
That accounts for the small fluctuations of the entropy in this range,
visible in Fig.~\ref{fig:zero-energy-entropy-average}.
In addition, as can be seen in the inset of Fig.~\ref{fig:min_energy}, the
zero energy states for $f<0.156$ are gapped.
%%%%%%%%%%%%%%%%%%%%%%%%%%%%%%
\item[(3)]
the bulk of the zero energy states lie in the interval $0.156\lesssim f \lesssim 1/5$.
In this interval the majority of the clusters have degenerate zero energy states
in every momentum sector. Let $(f_0, S_\maxx)$ be the maximum point of 
$S_0(f)$ function; our results indicate that $f_0\approx 0.178$ and $S_\maxx\approx 0.152$.  
Van Eerten~\cite{vanEerten:2005}, using a  transfer matrix, found a numerical 
bound  on the Witten index, which corresponds~\cite{Huijse:2011vn} to $S_\maxx \ge 0.13$,
and also indicates $f_0 \approx 0.18$, with which our results are consistent.
%%%%%%%%%%%%%%%%%%%%%%%%%%%%%%
\item[(4)]
Finally, at the upper end of the interval of zero states in the vicinity of $f\simeq1/5$, 
the slope of the entropy function $S_0(f)$ appears to become much steeper.
Thus we cannot rule out the possibility that the zero-energy state interval may end below or at $f=1/5$ 
in the thermodynamic limit.
\end{itemize}
%%%%%%%%%%%%%%%%%%%%%%%%%%%%%%

We have also made a study of the energy gaps in this filling
interval in clusters that have zero states, i.e. the smallest
non-zero eigen-energy (from any momentum sector).
The main prior result about gaps in this model on the
triangular lattice is that ladders with $f=1/6$ are gapless~\cite{Huijse:2011vn}.

As just noted, zero-energy states with $f<0.156$ appear
to be gapped.  In the filling interval $(0.156,1/5)$ we attempted
to study the gaps by averaging over different clusters, in the
same spirit as Eq.~\eqref{eq:entropy}. However, although
we see clear trends, the behavior of the gaps depends on
the clusters in an apparently irregular way, so the results
are trustworthy only within an order of magnitude.
Since the gaps are strongly (but not always monotonically)
decreasing with $N$, we only average within groups of clusters
having the same $(F,N)$.

Within our two-dimensional clusters having typically $\sim 40$ sites,
there seemed to be particularly low gaps around two fillings
$f\gtrsim0.170$ and $f\approx 0.190$, with a maximum 
around $f\approx 0.179$.  It appears, in fact, that these
two minima represent a supersymmetric pair with $F$ and $F+1$
fermions, e.g. fillings 7/42 and 8/42, and we conjecture that
in the thermodynamic limit there is one minimum occuring
at $f\approx 0.178$, where the entropy function is highest.
The gap values at the minimum decay at least as fast 
as $1/2^F$ (up to $F=8$ where our data is complete), and
are $10^{-3}$--$10^{-1}$ in our largest systems.  
At other fillings around the middle of the interval 
$(0.156,1/5)$ we still see a decaying trend with F, but slower. 
For $f\lesssim 0.163$ and $f\gtrsim 0.195$ gaps larger
than 0.1 are seen even in our largest systems, so we cannot
definitely say whether these are gapped.

\subsection{Phase diagram at filling $f\lesssim 1/3$}
\label{sec:fto1:3}

 It is generically expected that strongly interacting lattice fermion models at high densities 
 phase separate into a high-density insulator,  such as the inert
$f=1/3$ crystal in Figure \ref{fig:typical}(f),
and a low-density liquid phase in which kinetic energy is dominant.
%% In the non-supersymmetric model 
%% with nearest-neighbor exclusion~\cite{Zhang:2003cr,Henley:2001dq} 
%% such a coexistence is seen in the square lattice case.
On the triangular lattice, we would guess that the coexisting liquid density 
falls around $f=0.21$ where the Hilbert space is largest.
(We used the Pauling approximation for the entropy of allowed
configurations, following the Appendix  of Ref.~\onlinecite{Zhang:2003cr}.)
That is roughly the filling at which the dense liquid is becoming
congested, hence less favorable, due to the hardcore constraint.

Just which dense state does this dense liquid coexist with?  
Prior research~\cite{Zhang:2003cr,Henley:2001dq} on the non-supersymmetric model on the square lattice showed that the
dense phase is \emph{not} the maximally occupied crystal.  Instead, it is
a crystal containing a dilute array of quantum-fluctuating strings,
which we call ``stripe-walls'' since each has a deficit of 
fermion density and is a domain wall of the crystal order~\footnote
{In Refs.~\onlinecite{Henley:2001dq} and \onlinecite{Zhang:2003cr},
they were called ``stripes''.}.  There is no previous literature
on such stripe-walls in the case of the triangular lattice for
\emph{fermions}, only for hardcore \emph{bosons} (in the context of 
$^4$He adsorbed on carbon nanotubes~\cite{Green:2000,Green:2002}.
Nevertheless, it was shown in Ref.~\onlinecite{Zhang:2003cr} that in an isolated 
stripe-wall, hardcore constraints do not allow particle exchanges,
hence the fermion and boson cases should be in fact equivalent.

The stripe-wall runs perpendicular to one of the bond directions;
fermions on the edge of a domain are free to hop in that direction,
owing to the relative shift of the other domain.  The deficit of
electrons is $1/3$ per step along the stripe-wall. Let the energy
per step be $E_\SW$: then the chemical potential associated with
(non-interacting) stripe-walls is $\mu_\SW=3 E_\SW$.  Thus, a
phase of dilute stripe walls is represented by an energy function
$E(f)= 1/3 - \mu_\SW (1/3-f)$.  $E(1/3)$=1/3 in the maximally
filled crystal, since there is no kinetic energy and 
the potential energy in \eqref{eq:model} is $F=N/3$.  

Note that as $f$ continues to decrease and the array becomes 
less dilute, collisions between adjacent stripe-walls 
(due to their quantum fluctuations) start to become significant.
This will typically reduce the kinetic energy, 
causing the $E(f)$ curve to bend upwards.  Thus, if $\mu_\SW$
exceeds the slope of a coexistence curve connecting to the
$f=1/3$ crystal, the stripe-wall array is stable at $f\lesssim 1/3$.
Its stability ends at the point where a tangent to the $E(f)$ curve
can be drawn to the coexisting liquid phase.
(Such a point should exist since $E(f)$ is upwards curving.)
By contrast, if $\mu_\SW$ is less than the slope of the
coexistence curve, the $f=1/3$ crystalline phase 
coexists directly with a hole-rich phase having $f \approx 0.22$. 
In the latter case, in the grand canonical-like ensemble, 
a first-order transition would be seen between the hole-rich  phase 
and the $f=1/3$ crystal.
To check which scenario holds in the supersymmetric model,
we must calculate the energy $E_\SW$.
 
Past studies~\cite{Henley:2001dq,Green:2000} 
showed that a single stripe-wall can be mapped exactly
to a one-dimensional chain 
with \emph{noninteracting} spinless fermions 
at half filling, with their hopping amplitude equal to $t=1$.
This mapping remains valid in the supersymmetric case, since all the accessible 
configurations for the stripe-wall have equal (and maximal) potential energy $F$.
Thus, the kinetic energy per step is $-2/\pi$, or $-6/\pi$ per removed
fermion; the potential energy is $+1$ per removed fermion.
Hence, we obtain $\mu_\SW=(1+6/\pi)= 2.91$.

For comparison, our numerics from exact diagonalization of the 
supersymmetric model showed that in the interval below $f=1/3$,
the energy is linear as a function of filling with a form 
$E(f)=\frac{1}{3}+\lambda \left(f-\frac{1}{3}\right)$, with 
$\lambda=2.92$, represented by the solid line (cf. Fig.~\ref{fig:min_energy}). 
Since we find that $\lambda\cong \mu_\SW$, this leaves undecided whether
a stripe-wall array is stable or the $f=1/3$ crystal
coexists directly with $f \approx 0.22$. 

The straight line fitting $E(f)$ would pass through $f\approx 0.22$,
close to the maximum filling of special zero-energy states we found above
$f=0.2$.  However, these states do not appear in 
Figure~\ref{fig:min_energy}, which is limited to robustly two-dimensional
clusters; the numerical $E(f)$ curve curves upwards in the vicinity of $f=0.22$
and appears to hit the $E=0$ axis at $f=0.20$--$0.205$.  That means
the zero-energy states have a smooth transition to $E>0$ states,
which are stable in a short interval up to $f=0.22$, and then
possibly coexisting with either a stripe-array or the $f=1/3$ crystal.

\section{Zero-energy states at maximum and minimum fillings}
\label{sec:zero-endpoints}

The entropy of zero states vanishes, according to Figure ~\ref{fig:zero-energy-entropy-average},
at filling 1/7 and slightly over filling 1/5.  Around those fillings, if 
zero states exist at all, they tend to be unique in their momentum sector and have
other special properties, including a tendency to crystal-like spatial orderings.

%%%%%%%%%%%%%%%%%%%%%%%%%%%%%%%%%%%%%%%%%%
%%% New Table I.  f=1/7 states
\begin{table}
\centering
\begin{tabular}{|c|c|c|c|}
\hline 
$N$ & $n$ & $\bfu$ & $\bfv$\tabularnewline
\hline 
\hline 
28 & 4 & $\left(6,2\right)$ & $\left(1,5\right)$\tabularnewline
\hline 
28 & 4 & $\left(6,4\right)$ & $\left(2,6\right)$\tabularnewline
\hline 
28 & 4 & $\left(10,2\right)$ & $\left(1,3\right)$\tabularnewline
\hline 
35 & 5 & $\left(8,5\right)$ & $\left(1,5\right)$\tabularnewline
\hline 
35 & 5 & $\left(13,4\right)$ & $\left(1,3\right)$\tabularnewline
\hline 
42 & 6 & $\left(8,3\right)$ & $\left(2,6\right)$\tabularnewline
\hline 
42 & 6 & $\left(7,0\right)$ & $\left(2,6\right)$\tabularnewline
\hline 
42 & 6 & $\left(9,3\right)$ & $\left(1,5\right)$\tabularnewline
\hline 
42 & 6 & $\left(15,3\right)$ & $\left(1,3\right)$\tabularnewline
\hline 
%%%%%%%%%%%%%%%%%%%
\end{tabular} \hspace{1cm} \begin{tabular}{|c|c|c|c|}
\hline 
$N$ & $n$ & $\bfu$ & $\bfv$\tabularnewline
\hline 
\hline 
49 & 7 & $\left(11,6\right)$ & $\left(1,5\right)$\tabularnewline
\hline 
49 & 7 & $\left(18,5\right)$ & $\left(1,3\right)$\tabularnewline
\hline 
49 & 7 & $\left(7,7\right)$ & $\left(0,7\right)$\tabularnewline
\hline 
49 & 7 & $\left(8,5\right)$ & $\left(3,8\right)$\tabularnewline
\hline 
56 & 8 & $\left(10,2\right)$ & $\left(2,6\right)$\tabularnewline
\hline 
56 & 8 & $\left(11,5\right)$ & $\left(2,6\right)$\tabularnewline
\hline 
56 & 8 & $\left(12,4\right)$ & $\left(1,5\right)$\tabularnewline
\hline 
56 & 8 & $\left(20,4\right)$ & $\left(1,3\right)$\tabularnewline
\hline 
56 & 8 & $\left(8,5\right)$ & $\left(0,7\right)$\tabularnewline
\hline 
%%%%%%%%%%%%%%%%%%%
\end{tabular}

\caption{The clusters that have a state at filling $f=1/7$ for $28\leq N\leq 56$. All the clusters
are commensurate to the $\sqrt 7 \times\sqrt 7 $ crystalline structure shown in Fig.~\ref{fig:typical}(b). 
In all cases there is a unique zero-energy state in 7 different momentum sectors with a leading configuration 
which resembles Fig.~\ref{fig:typical}(b).
\label{tab:low_filling_clusters}}
\end{table}
%%%%%%%%%%%%%%%%%%%%%%%%%%%%%%%%%%%%%%%%%%

\begin{figure}
\tikzinput{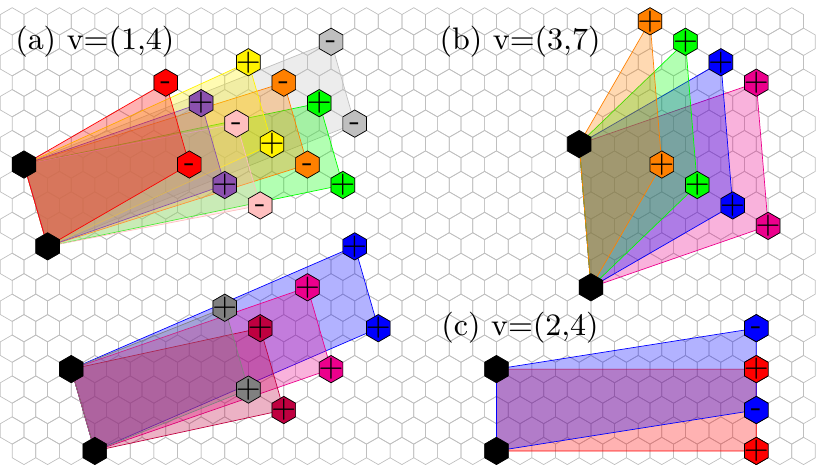}
\caption{ 
\textbf{A compilation of clusters with zero-energy states at filling $f>\frac{1}{5}$} for (a) $\bfv=(1,4)$, (b) $\bfv=(3,7)$ and (c) $\bfv=(2,4)$.
For each case, we mark with different colors the vectors $\bfu$ of all the clusters with a high filling zero-energy state. The symmetrically equivalent $\bfu$'s use the same color (color online). The $+$ and $-$ signs correspond to 
center of mass momentum ${\bfK}=(0,0)$ and ${\bf}=(0,\pi)$ respectively.
\label{fig:high_fill_clusters}}
\end{figure}

\subsection{Zero-energy states at filling $f\sim 1/7$}
\label{sec:f=1:7}

The zero energy states near the lower end of the interval 
$[f_-,f_+]$ have been discussed in Ref.~\onlinecite{Huijse:2011vn} but apparently not in detail. The authors, numerically, identified only one cluster, 
$(7,0)\times (0,7)$, having a zero energy state at $f=1/7$. 
In contrast, we found striking behaviors exactly at $f=1/7$ for a variety of clusters.

In Table ~\ref{tab:low_filling_clusters} we present all the clusters with zero-energy states
at $f=1/7$. Interestingly, those are all of the 
%% symmetrically inequivalent 
distinct clusters which are commensurate 
with the $\sqrt 7 \times \sqrt 7$ crystalline order shown in 
Fig.~\ref{fig:typical}(b), i.e.  both the vectors $\bfu$ and $\bfv$ 
are lattice vectors of that pattern.
The leading configurations of that wave function consist 
of the seven possible translations of the same perfectly ordered
pattern.

\begin{figure}
\subfigure[]{\tikzinput{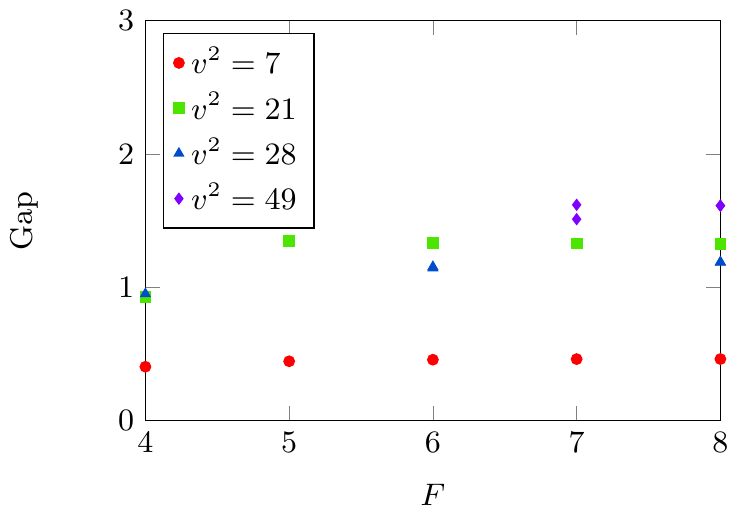}\label{fig:gap17}}
\subfigure[]{\tikzinput{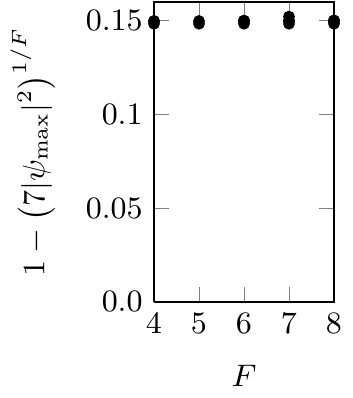}\label{fig:psimax}}\subfigure[]{\tikzinput{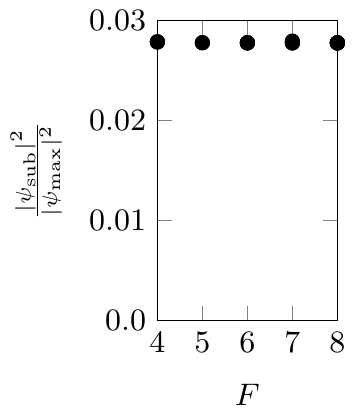}\label{fig:psisub}}
\caption{
Nature of the zero-energy states listed in 
Table~\ref{tab:low_filling_clusters}.
(a) The gap of the unique zero-energy state at $f=1/7$ as a function of the number of particles $F$.
The 3-leg ladders with $v=(1,3)$ approach a different value for large $F$ than the
rest of the clusters. The geometry of the ladders is such that each unit cell of the $\sqrt 7 \times \sqrt 7$
crystalline order wraps around it self. (b)  
The quantity $q=1- (7 |\psi_{\maxx}|^2)^{1/F} $, 
the inferred probability per crystal site to
find its fermion one site away, is independent of system size,
as expected for long-ranged crystalline order. 
(c) The weight of a sub-leading configuration (in which one
fermion is off its site)  relative to a leading one. 
This is also independent of $F$ and 
approximately equal to $\frac{1}{6}q/(1-q)$, as expected.
\label{fig:data17}}
\end{figure}

We find that those zero-energy states are gapped. 
Shown in Fig.~\ref{fig:gap17} is the gap plotted against the cluster size,
grouped as strips of different widths.
We notice that the strip of shortest width, 
$\bfv=(1,3)$ [i.e. three-leg ladders],
 seems to have a gap approaching $0.46$,
whereas the strips with longer $\bfv$ 
have gaps larger than 1. 
We remark that the three-leg ladders are not representative
 of the thermodynamic limit, since the nearest neighbors 
transverse to the strip are images of the  same fermion.

It is possible to understand the crystalline order within a very simple picture,
based on an eigenstate in which the symmetry is already broken.
Our main hypothesis is that, to a good approximation, the probability
$|\psi|^2$, as a function of configuration, is a direct product
of independent wavefunctions of each fermion hopping in a separate hexagon of
7 sites, centered on one of the ideal crystal sites.  Within each hexagon
the weight is $(1-q)$ on the center site and $q/6$ on the other sites.  
The leading configuration is when every fermion is on a center site and the
leading weight is $|\psi_\maxx|^2 = (1-q)^F/7$, where the factor of 7 signifies that
the actual eigenstate in any momentum sector is a superposition
of seven shifted copies of that eigenstate, which have no
configurations in common. Indeed in Fig.~\ref{fig:psimax}
we plot  $q=1-\left( 7 |\psi_{\maxx}|^2 \right) ^{1/F}$ and we find it independent
of $F$ with a limiting value of $q\approx0.148$.

A sub-leading term in the many-fermion wavefunction
(as we verified by inspecting the wavefunction) goes with a configuration
in which each of $(F-1)$ fermions is on the center site of its hexagon,
but one fermion is on a neighbor site in its hexagon.
Hence the sub-leading amplitude should be $|\psi_\subb|^2 = \frac{1}{7}\frac{q}{6}(1-q)^{F-1}$
and $|\psi_\subb|^2/|\psi_\maxx|^2 = q/6(1-q)$.  
According to Fig.~\ref{fig:psisub}, the limiting value of this ratio
is 0.028 which corresponds to $q\approx0.144$. 
The good agreement of  this to the $q$ value mentioned above
supports the validity of our simple picture. 

\begin{figure}
\subfigure[$(8,4),(2,6)$, $f=\frac{6}{40}$]{\tikzinput{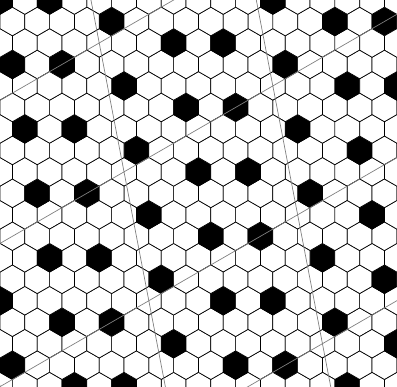}}
\subfigure[$(7,3),(1,6)$, $f=\frac{6}{39}$]{\tikzinput{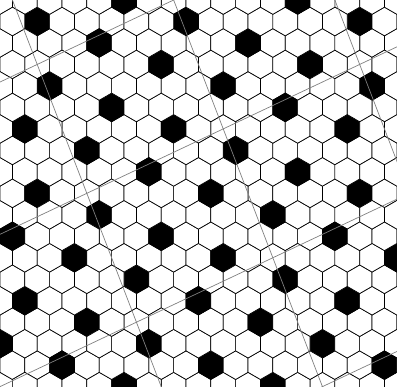}}
\caption{
The leading configurations for two particular clusters in the interval $1/7<f<0.156$. Both correspond
to unique and gapped zero energy states at a particular momentum sector. Both of them
have the form of stripes of the $f=1/7$ crystal separated by domain walls.
}\label{fig:f>1/7}
\end{figure}

The character of the $f=1/7$ zero energy ground states survives in a small interval, $1/7<f<0.156$
in which only some of the clusters have zero energy states, 
in only a few momentum sectors.
Similarly to the $f=1/7$ case, those states are found to be unique 
and protected by a gap which gets smaller as the $f$ of these states gets larger.
The inset of Fig.~\ref{fig:min_energy} shows that the energy
vs filling does exhibit a gap in this interval.
Looking at the structure of the wave function 
reveals that the leading configuration (and the symmetrically equivalent ones) constitute an $f=1/7$ pattern
with a domain wall as shown in Fig.~\ref{fig:f>1/7} whereas 
the subleading configurations are those with one fermion hopped out of this
pattern.

\subsection{Zero-energy states at filling $f\gtrsim 1/5$}
\label{sec:f=1:5}

The existence of zero-energy states for filling $f > 1/5$ has been
a subject of debate.
Jonsson in a recent study~\cite{Jonsson:2010kx} analyzed a collection of homology cycles on the triangular grid and showed that for a specific class of configurations there is an upper bound on the existence of zero-energy ground states which is $f =1/5$. Moreover, Huijse~\etal~\cite{Huijse:2011vn} through extensive exact diagonalization studies and analytical arguments found agreement of the Jonsson conjecture with numerical results on the triangular grid. 
Our studies have found zero-energy states beyond the filling $f=1/5$.

\subsubsection{Occurrence and excitation gaps}

Figure \ref{fig:high_fill_clusters}
displays all the clusters where $f> 1/5$ zero-energy states are observed. 
Such states are observed in both elongated and roughly isotropic cluster shapes, 
suggesting that their existence is not an artifact of the aspect ratio
(but it could be finite-size related). 
We find that for a few clusters there are unique zero-energy
states at a single momentum sector which is either $\left(0,0\right)$
or $\left(0,\pi\right)$ depending on the cluster.  
Table \ref{tab:high_capacity_clusters}
shows all the clusters and momentum sectors with $f>1/5$
zero-energy states. 

%%%%%%%%%%%%%%%%%%%%%%%%%%%%%%%%%%%%%%%%%%%%%%%%%%%%%%%%%%%%%%%%%%%%%%
%%%% New Table II (f>0.2)
\begin{table}
\centering
\begin{tabular}{|c|c|c|c|c|c|c|c|c|}
\hline 
$N$ & $F$ & $f$ & $\bfu$ & $\bfv$ & $\bfK$ & $N_\lead$ & $W_\lead$ & $\Delta E$ \tabularnewline
\hline 
\hline 
28 & 6 & 0.2143 & $\left(8,4\right)$ & $\left(1,4\right)$ & $\left(0,\pi\right)$ & 9 & 0.6667 & 1.1038 \tabularnewline
\hline 
33 & 7 & 0.2121 & $\left(9,3\right)$ & $\left(1,4\right)$ & $\left(0,0\right)$ & 12 & 0.4893 & 0.9327 \tabularnewline
\hline 
38 & 8 & 0.2105 & $\left(10,2\right)$ & $\left(1,4\right)$ & $\left(0,\pi\right)$ & 16 & 0.3590 & 0.8290 \tabularnewline
\hline 
43 & 9 & 0.2093 & $\left(12,5\right)$ & $\left(1,4\right)$ & $\left(0,0\right)$ & 20 & 0.2475 & 0.7068 \tabularnewline
\hline 
48 & 10 & 0.2083 & $\left(13,4\right)$ & $\left(1,4\right)$ & $\left(0,\pi\right)$ & 25 & 0.1705 & 0.5978 \tabularnewline
\hline 
53 & 11 & 0.2075 & $\left(14,3\right)$ & $\left(1,4\right)$ & $\left(0,0\right)$ & 30 & 0.1127 & 0.5075 \tabularnewline
\hline 
58 & 12 & 0.2069 & $\left(16,6\right)$ & $\left(1,4\right)$ & $\left(0,\pi\right)$ & 36 & 0.0745 & 0.4364 \tabularnewline
\hline 
\hline 
29 & 6 & 0.2069 & $\left(8,3\right)$ & $\left(1,4\right)$ & $\left(0,0\right)$ & 2 & 0.3252 & 0.2407 \tabularnewline
\hline 
34 & 7 & 0.2059 & $\left(9,2\right)$ & $\left(1,4\right)$ & $\left(0,0\right)$ & 16 & 0.4028 & 0.6079  \tabularnewline
\hline 
44 & 9 & 0.2093 & $\left(12,4\right)$ & $\left(1,4\right)$ & $\left(0,0\right)$ & 40 & 0.2127 & 0.3289\tabularnewline
\hline 
54 & 11 & 0.2037 & $\left(15,6\right)$ & $\left(1,4\right)$ & $\left(0,0\right)$ & 80 & 0.1001 & 0.3097 \tabularnewline
\hline 
\hline 
44 & 9 & 0.2093 & $\left(11,0\right)$ & $\left(2,4\right)$ & $\left(0,0\right)$ & 8 & 0.1337 & 0.7345\tabularnewline
\hline 
44 & 9 & 0.2093 & $\left(12,2\right)$ & $\left(2,4\right)$ & $\left(0,\pi\right)$ & 8 & 0.1337 & 0.6338 \tabularnewline
\hline 
\hline 
24 & 5 & 0.2083 & $\left(6,6\right)$ & $\left(3,7\right)$ & $\left(0,0\right)$ & 4 & 0.6379 &0.9461 \tabularnewline
\hline 
34 & 7 & 0.2059 & $\left(7,5\right)$ & $\left(3,7\right)$ & $\left(0,0\right)$ & 14 & 0.6198 & 0.6386 \tabularnewline
\hline 
44 & 9 & 0.2093 & $\left(8,4\right)$ & $\left(3,7\right)$ & $\left(0,0\right)$ & 32 & 0.4144 & 0.5823 \tabularnewline
\hline 
54 & 11 & 0.2037 & $\left(9,3\right)$ & $\left(3,7\right)$ & $\left(0,0\right)$ & 66 & 0.2531 & 0.7621 \tabularnewline
\hline 
\end{tabular}

\caption{Clusters in which a high filling zero-energy state is found.
$N$ is the number of sites, $F$ the number of fermions, and $f=F/N$ is the filling;
the vectors $\bfu$ and $\bfv$ define the cluster unit cell.
In each case, there is a unique zero-energy eigenfunction in momentum sector $\bfK$.
$N_\lead$ is the number of terms in that (normalized) wavefunction having 
identical maximum amplitude $|\psi_\lead|$ and not equivalent by translations;
 $W_\lead=N_\lead|\psi_{\lead}|^2$ is the combined weight of such terms;
it can be seen that $W_\lead$ tends to vanish with increasing system size.
Also, $\Delta E$ is the gap to the next state in the same sector
(of $F$ and $\bfK$)}
\label{tab:high_capacity_clusters}
\end{table}
%%%%%%%%%%%%%%%%%%%%%%%%%%%%%%%%%%%%%%%%%%%%%%%%%%%%%%%%%%%%%%%%%%%%%%

The shapes of cluster that tend to support
zero-energy states with $f\gtrsim 1/5$ are
shown in Table~\ref{tab:high_capacity_clusters}.
As illustrated in Figure~\ref{fig:high_fill_clusters},
these clusters fall into sequences,
each having fillings that approach $f=1/5$ from above
in the thermodynamic limit.

The clusters with $\bfv=\left(1,4\right)$ in the beginning rows
of Table~\ref{tab:high_capacity_clusters} fall in a sequence
with $\bfu=\left(0,2-\epsilon\right)+m\left(1,-1\right)$ modulo $\bfv$
for $\epsilon=0,1$ and $m=6,7, \ldots, ,12$. These have $F=m$ fermions
in $N=5m+\epsilon-2$ sites so the filling is $f=/[5+(\epsilon-2)/m]$. 
The momentum
where the zero-energy states appear is $\bfK=\left(0,0\right)$, 
except when $\epsilon=0$ and $m$ is even we have $\bfK=\left(0,\pi\right)$.
(The $\epsilon=1$ zero-energy states appear only for $m=6,7,9,11$,
so they may be a finite size effect.)
Also, the last rows of the table are four clusters 
having $\bfv=\left(3,7\right)$ and the other vector in a sequence
$\bfu=4\left(1,2\right)+ m \left(1,-1\right)$ with $m=2, \ldots$.
(the $m=1$ cluster was omitted from the table because $N<25$.)
These have $F=2m+1$ fermions in $N= 10m-1$ sites, for a filling
which is obviously greater than $1/5$.

The gaps of the $\bfv=(1,4)$ ladders shown in table~\ref{tab:high_capacity_clusters}
can be fitted to a decaying exponential function of the system size, which extrapolates to 
a gapless state in the thermodynamic limit. Indeed, within this series the gaps fit well
to $\exp(-\textrm{Const} N^2)$. We know (see Section~\ref{sec:f=1:7}) that in a 
crystal the amplitudes should decay exponentially, so this is suggestive
of a liquid phase, or perhaps a crystal with unbounded fluctuations,
similar to a classical crystal in $d=2$ at nonzero temperature.

The clusters at width $\bfv=(3,7)$, the only ones in this table that are plausibly two-dimensional, have the largest gaps --
of order unity --  suggesting the presence of a gapped state in this filling.  
For the $\bfv=(2,4)$ ladders  the data are insufficient to draw a conclusion.

\subsubsection{Density correlations and coincident wavefunction amplitudes}

\begin{figure}
%%%%%%%%%%%%%%%%%%%%%%%%%%%%%%%%
\subfigure[]{\tikzinput{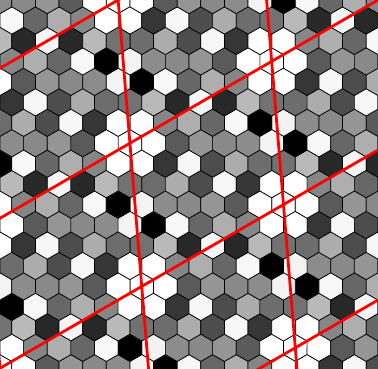}}\label{fig:correlation_8_4_3_7}
\subfigure[]{\tikzinput{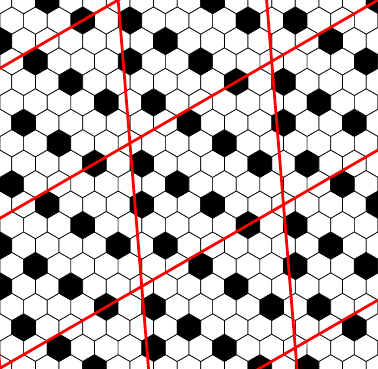}}\label{fig:configuration_8_4_3_7}

\subfigure[]{\tikzinput{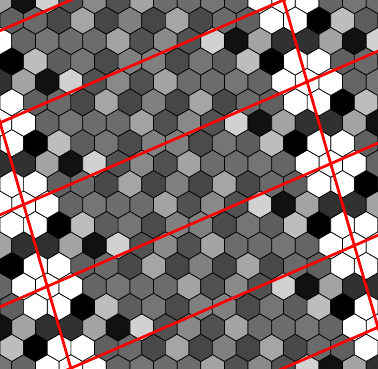}}\label{fig:correlation_15_6_1_4}
\subfigure[]{\tikzinput{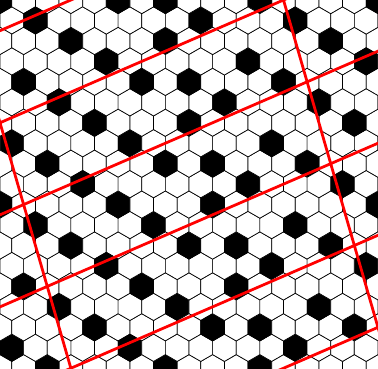}}\label{fig:configuration_15_6_1_4}

\subfigure[]{\tikzinput{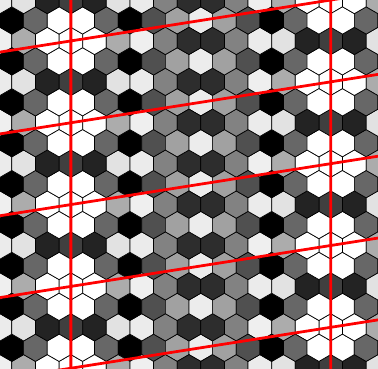}}\label{fig:correlation_12_2_2_4}
\subfigure[]{\tikzinput{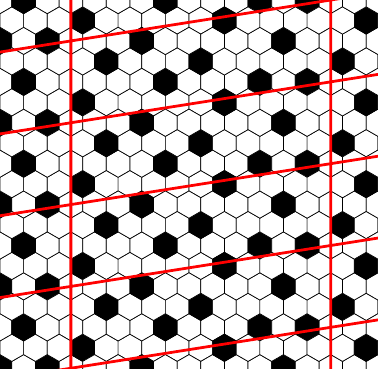}}\label{fig:configuration_12_2_2_4}
%%%%%%%%%%%%%%%%%%%%%%%%%%%%%%%%%%%
\caption{
The density density correlation (left panels) and the leading configuration (right panels) for the unique
zero-energy states at $f\gtrsim1/5$. 
In each case, the unit cells are marked by light dotted lines,
starting from the left lower corner.
note in (c,d) the second set of $(1,4)$ edges does not pass
through the panel, while in (e,f) the second
set of $(2,4)$ edges is the panel's left boundary.
(a), (b) $\left(8,4\right),\left(3,7\right)$ at $f=9/44$, 
(c), (d) $\left(15,6\right),\left(1,4\right)$ at $f=11/54$,
(e), (f) $\left(12,2\right),\left(2,4\right)$ at $f=9/44$.
(a), (c) and (d): map of the density density correlation $\langle\hat \rho(\bfr)\hat\rho(\bfr + \bfR)\rangle$
as a function $R$. In each panel many copies of the unit cell (marked by thin lines) are shown;  
darker grays indicate higher correlations, except that the origin is made white for better contrast. 
At certain separations the correlation function is almost zero, but its only exact zeroes are the nearest neighbors
to the origin (and points equivalent by translation), which are excluded by the reference fermion.
(b), (d) and (f): one of the leading configurations all having identical weights as shown in table. \ref{tab:high_capacity_clusters}.
Each configuration has the covering property, i.e. every unoccupied site is neighbor to at least one fermion, and
consequently the potential energy has a the minimum value.}
\label{fig:density_configurations}
\end{figure}

To further understand the ground state wavefunctions at fillings $f>1/5$, 
we examined the density correlation functions in these wavefunctions
[Fig.~\ref{fig:density_configurations}(a,c,e)].
The density-density correlation function in each of these states
suggests a non-trivial order, which in some cases it is stripe-like, 
whereas in others it corresponds to a more isotropic structure [cf. Fig. \ref{fig:density_configurations} (a,c,e).]

We also examined the leading amplitude configurations in each
wavefunction
[examples in Fig.~\ref{fig:density_configurations}(b,d,f)].
The combined weight of those leading configurations appears 
to vanish with the system size, but that happens 
even when the leading configurations strongly dominate
the behavior, e.g. in the $f=1/7$ crystal 
[see Figure~\ref{fig:data17}(b)] where they decay
as $\exp(-{\rm const} N$.
For the $f\gtrsim 1/5$ zero-energy wavefunctions, the decay
appears to be faster, perhaps as s $\exp(-{\rm const} N^2$.
(This can be inferred from the last column of 
Table~\ref{tab:high_capacity_clusters}.)
This indicates that the $f\gtrsim 1/5$ states are less ordered
than the $f=1/7$ crystal.

\subsubsection{``Magic'' wavefunctions} 
\label{sec:magic}

All of the $f\gtrsim 1/5$ zero-energy wavefunctions have
the unusual property of ``coincident amplitudes'',
meaning there are multiple
configurations having the identical amplitudes (apart from sign).
We have checked this only for those of maximum 
amplitude, i.e. the ``leading'' configurations.
In all cases we know, these leading configurations all have the
minimum possible potential energy $V_\minn$.

The coincident-amplitude property is somewhat reminiscent of the 
Rokhsar-Kivelson (RK) eigenstates~\cite{Rokhsar:1988},
in which \emph{all} configurations have equal weight.
Generalizations of the 
RK construction~\cite{Ardonne:2004uq,Henley:2004ys,Papanikolaou:2007kx, Castelnovo:2005cr}
might inspire future possibilities of analytic approaches
to the supersymmetric model:  e.g. one might construct the \emph{exact} 
ground state wavefunction for the given cluster; or possibly, the
quantum ground state might have the same weights as the
Boltzmann ensemble of some classical model, 
which would assist in understanding the 
degree and nature of order in the thermodynamic limit.

In the present work, we could only recognize the coincident-amplitude 
property when the zero state is unique/non-degenerate in its momentum sector; 
otherwise, the condition is ill-defined, since the amplitudes depend on which 
linear combination of the degenerate states is provided by the Lanczos solver.
Thus, such wavefunctions are likely to be identified only
near the endpoints of the interval of
fillings with zero energy states, i.e. around
$f\approx 0.15$ or $f\approx 0.20$; in practice we only
identified them for $f\gtrsim 0.20$.  

Furthermore, the leading configurations in 
Fig. \ref{fig:density_configurations}(b) and (d) are dominated by
fragments of the $f=1/5$ crystal from Fig.~\ref{fig:typical}(e),
which is the densest simple crystal in which the fermions have
available hoppings and the potential energy has the minimum 
value of $F$.  In the actual leading configurations, some
triangles from the other crystals in Fig.~\ref{fig:typical}
get mixed in, which allows more quantum fluctuations and
lowers the kinetic energy.  
Notice these are very anisotropic structures, in that almost
all allowed hoppings are in the same directions (aligned with
the long direction of the $f=1/5$ triangles).
The configuration in Fig. \ref{fig:density_configurations}(f) appears
to be dominated instead by rows of the $f=1/6$ cell from 
Fig.~\ref{fig:typical}(d) which explains why the $(12,2),(2,4)$ cluster
is more isotropic than the other clusters.

\section{Concluding remarks}
\label{sec:conclusions}

The exact diagonalization study we performed on a variety of finite clusters
illuminated certain aspects of the phase diagram of the supersymmetric
lattice model of Eq.~\ref{eq:model}.
The main findings are the presence of gapped crystalline zero energy
states for $f\sim1/7$,
gapless quasi-ordered states for $f>1/5$ in violation of Jonsson's
conjecture
and a zero state entropy which peaks at $f\sim 0.178$.
Our results are not conclusive about the exact phase diagram in several
places and we now speculate on the possible scenarios, in light of our
results.

Our finding of an energy gap at $f=1/7$,  as shown in
Section~\ref{sec:f=1:7}, implies that the Fermi liquid phase cannot
connect continuously to the zero state at $f=1/7$. Either there is a
first-order transition, so that coexistence of the
Fermi liquid with the $f=1/7$ crystal is present in a small interval
below $f=1/7$, or else the crystal phase extends (with $E>0$) to
lower fillings.  A specific mechanism for the latter alternative is
that the crystal is doped with fermion vacancies (which thus form a
Fermi liquid within the crystal), which destabilizes (i.e. melts) the
crystal at a sufficient density of vacancies.  Additional studies will
be needed to decide between these scenarios.

The next open question is the behavior immediately above $f=1/7$.
Our results suggest there is a family of states made by
combining strips of the $f=1/7$ and $f=1/6$ crystals:
each such state is a zero-energy state similar to the 
$f=1/7$ crystal in being gapped and in having a crystal-like symmetry breaking.
This would imply a dense set of rational fillings 
above $f = 1/7$ that are, in some sense, isolated from each 
other rather than forming a continuum.

We conjecture that the fillings in the interval $0.156<f<0.2$ are all
{\it gapless} in the thermodynamic limit,
based on the numerical observations discussed at the end of
section~\ref{sec:zero+entropy}. If this is true in the thermodynamic
limit, it would fit naturally with a liquid-like phase, in
which the wavefunction has a vast number of
important configurations, each of them mixing the
units from all the crystals shown in Figure~\ref{fig:typical},
and the important configurations are accessible to each
other by small steps corresponding to local rearrangements
of these units. This scenario could be reconciled with the {\it
apparent} increase of the gaps away from the center of that interval at
$f=0.178$, in the following speculative picture. Say that the entropy
(per site) of {\it low}-energy states has a similar dependence on $f$
as the entropy of {\it zero}-energy states.
Then, for $f$ near the endpoints and with a finite size $N$,
the expected number of low-energy states in a cluster might be
less than unity, meaning that sometimes the cluster has none.

For the clusters with zero energy states at $f\gtrsim 1/5$
we find non-degenerate zero-energy states in just a single momentum sector.
The filling of those states seems to converge to $1/5$ for
increasing system size; furthermore, we checked that if we
make a new cluster that simply doubles one of these clusters,
and keep the same $f$, the larger cluster does {\it not}
generically have a zero energy state.  It is unclear how
our results relate to the thermodynamic limit.  These states showed
more than one kind of ordering tendency, depending on the cluster
(as noted at end of Sec.~\ref{sec:f=1:5}), but the tendency
is that most of the available hoppings lie along
a particular axis.
This might be explained by a local order incorporating many
fragments of the $f=1/5$ crystal [Figure~\ref{fig:typical}(e)],
which only allows hops along one lattice axis.

The corresponding wave functions at $f\gtrsim 0.2$
are dominated by many 
configurations with equal (``coincident'') amplitudes 
that always belong to the lowest potential energy sector,
(Sec.~\ref{sec:magic}).
We certainly cannot rule out the possibility that similar wavefunctions
exist for other zero states, since we are only able to detect
this property when there is a unique zero state in a particular sector.
The regularities in these wavefunctions suggest
the possibility that some analytic structure,
even an exact solution, may be found for the
zero-energy ground states.

Finally, at both ends of the range (1/5,1/3), we naively expect some
form of coexistence of a hole-rich
phase with the $f=1/3$ inert crystal, or this crystal doped by
an array of stripe-walls.  The $E(f)$ curve shows a straight line
at very nearly the slope expected from a stripe-wall array.
However, the picture from our exact diagonalizations is not
quite consistent with any scenario:  the $E(f)$ curve due to an
array ought to curve upwards rather than be a straight line;
in contrast, in the case of coexistence a straight
tie-line is not expected in a finite system due to the additional
cost of the domain wall.  One intriguing possibility is that
the domain-wall cost is zero: that is,
the phase at $f\gtrsim 0.5$ and the stripe-wall array dissolve
into each other.  This is not implausible: the stripe-wall
array is an $f=1/3$ crystal with some units of the $f=1/5$
crystal appearing along the stripe-walls, and it has hoppings
entirely along one of the crystal axis directions.  Thus,
its properties are similar to some of the zero states we
observed at $f\gtrsim 1/5$, so perhaps the quasi-order we
found in those states is simply a generalization of the
stripe-wall kind of state. It is worth noting that in 
Ref.~\onlinecite{Huijse:2011vn}, it was argued that the ground state 
at $f=1/4$ is gapped. It would be interesting to identify the deep origin of this gap, given that
stripe-walls, as well as coexisting stripe-walls should be by all odds gapless.

Overall, a possible interpretation of our consistent finding of a strong
bias towards ordering in finite clusters, is that the supersymmetric
model is
a multicritical point in the parameter space of Hamiltonians, so that
many different
kinds of order may be stabilized by infinitesimal perturbations.
The various sizes and shapes of the system introduce similar biases, so that
different large clusters may be occupied by fragments of competing
ordered states which
are all valid and degenerate in the thermodynamic limit.
In future studies, it may be worth to add terms to the Hamiltonian
\eqref{eq:model} so as to explore the possible relation of supersymmetry
to multicriticality~\cite{Huijse:2008zr,Lee:2007oq}.

\acknowledgments{We acknowledge support from the Lee Kuan Yew fellowship (DG), 
NSF grant DMR-1005466 (CLH) and DTRA Grant No. 1-10-1-0021 (SP). 
We thank P. Fendley, E. Fradkin, L. Huijse, and R. Z. Lamberty  for discussions 
and for comments on the text.}

\bibliography{susy}{}

\begin{thebibliography}{27}
\expandafter\ifx\csname natexlab\endcsname\relax\def\natexlab#1{#1}\fi
\expandafter\ifx\csname bibnamefont\endcsname\relax
  \def\bibnamefont#1{#1}\fi
\expandafter\ifx\csname bibfnamefont\endcsname\relax
  \def\bibfnamefont#1{#1}\fi
\expandafter\ifx\csname citenamefont\endcsname\relax
  \def\citenamefont#1{#1}\fi
\expandafter\ifx\csname url\endcsname\relax
  \def\url#1{\texttt{#1}}\fi
\expandafter\ifx\csname urlprefix\endcsname\relax\def\urlprefix{URL }\fi
\providecommand{\bibinfo}[2]{#2}
\providecommand{\eprint}[2][]{\url{#2}}

\bibitem[{\citenamefont{Weinberg}(2005)}]{weinberg-book}
\bibinfo{author}{\bibfnamefont{S.}~\bibnamefont{Weinberg}},
  \emph{\bibinfo{title}{The {Q}uantum {T}heory of {F}ields, {V}olume 3:
  {S}upersymmetry}} (\bibinfo{publisher}{Cambridge Univ Press},
  \bibinfo{year}{2005}).

\bibitem[{\citenamefont{Fendley
  et~al.}(2003{\natexlab{a}})\citenamefont{Fendley, Schoutens, and
  De~Boer}}]{Fendley:2003uq}
\bibinfo{author}{\bibfnamefont{P.}~\bibnamefont{Fendley}},
  \bibinfo{author}{\bibfnamefont{K.}~\bibnamefont{Schoutens}},
  \bibnamefont{and} \bibinfo{author}{\bibfnamefont{J.}~\bibnamefont{De~Boer}},
  \bibinfo{journal}{Physical Review Letters} \textbf{\bibinfo{volume}{90}},
  \bibinfo{pages}{120402} (\bibinfo{year}{2003}{\natexlab{a}}).

\bibitem[{\citenamefont{Fendley
  et~al.}(2003{\natexlab{b}})\citenamefont{Fendley, Nienhuis, and
  Schoutens}}]{Fendley:2003fk}
\bibinfo{author}{\bibfnamefont{P.}~\bibnamefont{Fendley}},
  \bibinfo{author}{\bibfnamefont{B.}~\bibnamefont{Nienhuis}}, \bibnamefont{and}
  \bibinfo{author}{\bibfnamefont{K.}~\bibnamefont{Schoutens}},
  \bibinfo{journal}{Journal of Physics A: Mathematical and General}
  \textbf{\bibinfo{volume}{36}}, \bibinfo{pages}{12399}
  (\bibinfo{year}{2003}{\natexlab{b}}).

\bibitem[{\citenamefont{Ilieva et~al.}(2006)\citenamefont{Ilieva, Narnhofer,
  and Thirring}}]{Ilieva:2006qf}
\bibinfo{author}{\bibfnamefont{N.}~\bibnamefont{Ilieva}},
  \bibinfo{author}{\bibfnamefont{H.}~\bibnamefont{Narnhofer}},
  \bibnamefont{and} \bibinfo{author}{\bibfnamefont{W.}~\bibnamefont{Thirring}},
  \bibinfo{journal}{Fortschritte der Physik} \textbf{\bibinfo{volume}{54}},
  \bibinfo{pages}{124} (\bibinfo{year}{2006}).

\bibitem[{\citenamefont{Hagendorf and Fendley}(2011)}]{Hagendorf:2011bh}
\bibinfo{author}{\bibfnamefont{C.}~\bibnamefont{Hagendorf}} \bibnamefont{and}
  \bibinfo{author}{\bibfnamefont{P.}~\bibnamefont{Fendley}},
  \bibinfo{journal}{Journal of Statistical Physics} pp. \bibinfo{pages}{1--34}
  (\bibinfo{year}{2011}).

\bibitem[{\citenamefont{Huijse et~al.}(2008)\citenamefont{Huijse, Halverson,
  Fendley, and Schoutens}}]{Huijse:2008zr}
\bibinfo{author}{\bibfnamefont{L.}~\bibnamefont{Huijse}},
  \bibinfo{author}{\bibfnamefont{J.}~\bibnamefont{Halverson}},
  \bibinfo{author}{\bibfnamefont{P.}~\bibnamefont{Fendley}}, \bibnamefont{and}
  \bibinfo{author}{\bibfnamefont{K.}~\bibnamefont{Schoutens}},
  \bibinfo{journal}{Physical Review Letters} \textbf{\bibinfo{volume}{101}},
  \bibinfo{pages}{146406} (\bibinfo{year}{2008}).

\bibitem[{\citenamefont{Fendley and Schoutens}(2005)}]{Fendley:2005ys}
\bibinfo{author}{\bibfnamefont{P.}~\bibnamefont{Fendley}} \bibnamefont{and}
  \bibinfo{author}{\bibfnamefont{K.}~\bibnamefont{Schoutens}},
  \bibinfo{journal}{Physical Review Letters} \textbf{\bibinfo{volume}{95}},
  \bibinfo{pages}{046403} (\bibinfo{year}{2005}).

\bibitem[{\citenamefont{{Huijse} and {Swingle}}(2012)}]{Huijse:2012}
\bibinfo{author}{\bibfnamefont{L.}~\bibnamefont{{Huijse}}} \bibnamefont{and}
  \bibinfo{author}{\bibfnamefont{B.}~\bibnamefont{{Swingle}}},
  \bibinfo{journal}{ArXiv e-prints}  (\bibinfo{year}{2012}),
  \eprint{1202.2367}.

\bibitem[{\citenamefont{Fendley and Hagendorf}(2011)}]{Fendley:2011ly}
\bibinfo{author}{\bibfnamefont{P.}~\bibnamefont{Fendley}} \bibnamefont{and}
  \bibinfo{author}{\bibfnamefont{C.}~\bibnamefont{Hagendorf}},
  \bibinfo{journal}{Journal of Statistical Mechanics: Theory and Experiment}
  \textbf{\bibinfo{volume}{2011}}, \bibinfo{pages}{P02014}
  (\bibinfo{year}{2011}).

\bibitem[{\citenamefont{Huijse et~al.}(2011)\citenamefont{Huijse, Mehta, Moran,
  Schoutens, and Vala}}]{Huijse:2011vn}
\bibinfo{author}{\bibfnamefont{L.}~\bibnamefont{Huijse}},
  \bibinfo{author}{\bibfnamefont{D.}~\bibnamefont{Mehta}},
  \bibinfo{author}{\bibfnamefont{N.}~\bibnamefont{Moran}},
  \bibinfo{author}{\bibfnamefont{K.}~\bibnamefont{Schoutens}},
  \bibnamefont{and} \bibinfo{author}{\bibfnamefont{J.}~\bibnamefont{Vala}},
  \bibinfo{journal}{Arxiv preprint arXiv:1112.3314}  (\bibinfo{year}{2011}).

\bibitem[{\citenamefont{Jonsson}(2010)}]{Jonsson:2010kx}
\bibinfo{author}{\bibfnamefont{J.}~\bibnamefont{Jonsson}},
  \bibinfo{journal}{Discrete \& Computational Geometry}
  \textbf{\bibinfo{volume}{43}}, \bibinfo{pages}{927} (\bibinfo{year}{2010}).

\bibitem[{\citenamefont{Witten}(1982)}]{Witten:1982fk}
\bibinfo{author}{\bibfnamefont{E.}~\bibnamefont{Witten}},
  \bibinfo{journal}{Nuclear Physics B} \textbf{\bibinfo{volume}{202}},
  \bibinfo{pages}{253} (\bibinfo{year}{1982}).

\bibitem[{\citenamefont{Lehoucq et~al.}(1998)\citenamefont{Lehoucq, Sorensen,
  and Yang}}]{arpack}
\bibinfo{author}{\bibfnamefont{R.~B.} \bibnamefont{Lehoucq}},
  \bibinfo{author}{\bibfnamefont{D.~C.} \bibnamefont{Sorensen}},
  \bibnamefont{and} \bibinfo{author}{\bibfnamefont{C.}~\bibnamefont{Yang}}
  (\bibinfo{publisher}{Soc for Industrial \& Applied Math},
  \bibinfo{year}{1998}), ISBN \bibinfo{isbn}{0898714079}.

\bibitem[{\citenamefont{Zhang and Henley}(2003)}]{Zhang:2003cr}
\bibinfo{author}{\bibfnamefont{N.~G.} \bibnamefont{Zhang}} \bibnamefont{and}
  \bibinfo{author}{\bibfnamefont{C.~L.} \bibnamefont{Henley}},
  \bibinfo{journal}{Physical Review B} \textbf{\bibinfo{volume}{68}},
  \bibinfo{pages}{014506} (\bibinfo{year}{2003}).

\bibitem[{\citenamefont{Henley and Zhang}(2001)}]{Henley:2001dq}
\bibinfo{author}{\bibfnamefont{C.~L.} \bibnamefont{Henley}} \bibnamefont{and}
  \bibinfo{author}{\bibfnamefont{N.~G.} \bibnamefont{Zhang}},
  \bibinfo{journal}{Physical Review B} \textbf{\bibinfo{volume}{63}},
  \bibinfo{pages}{233107} (\bibinfo{year}{2001}).

\bibitem[{\citenamefont{Fradkin}(1991)}]{Fradkin:1991fk}
\bibinfo{author}{\bibfnamefont{E.}~\bibnamefont{Fradkin}},
  \emph{\bibinfo{title}{Field theories of condensed matter systems}}
  (\bibinfo{publisher}{Addison-Wesley Reading, Massachusetts},
  \bibinfo{year}{1991}).

\bibitem[{\citenamefont{Zhang and Henley}(2004)}]{Zhang:2004}
\bibinfo{author}{\bibfnamefont{N.~G.} \bibnamefont{Zhang}} \bibnamefont{and}
  \bibinfo{author}{\bibfnamefont{C.~L.} \bibnamefont{Henley}},
  \bibinfo{journal}{The European Physical Journal B - Condensed Matter and
  Complex Systems} \textbf{\bibinfo{volume}{38}}, \bibinfo{pages}{409}
  (\bibinfo{year}{2004}).

\bibitem[{\citenamefont{Fendley and Schoutens}(2009)}]{Fendley:2009bh}
\bibinfo{author}{\bibfnamefont{P.}~\bibnamefont{Fendley}} \bibnamefont{and}
  \bibinfo{author}{\bibfnamefont{K.}~\bibnamefont{Schoutens}},
  \bibinfo{journal}{New Trends in Mathematical Physics} pp.
  \bibinfo{pages}{277--284} (\bibinfo{year}{2009}).

\bibitem[{\citenamefont{van Eerten}(2005)}]{vanEerten:2005}
\bibinfo{author}{\bibfnamefont{H.}~\bibnamefont{van Eerten}},
  \bibinfo{journal}{Journal of Mathematical Physics}
  \textbf{\bibinfo{volume}{46}}, \bibinfo{eid}{123302}
  (pages~\bibinfo{numpages}{8}) (\bibinfo{year}{2005}).

\bibitem[{\citenamefont{Green and Chamon}(2000)}]{Green:2000}
\bibinfo{author}{\bibfnamefont{D.}~\bibnamefont{Green}} \bibnamefont{and}
  \bibinfo{author}{\bibfnamefont{C.}~\bibnamefont{Chamon}},
  \bibinfo{journal}{Phys. Rev. Lett.} \textbf{\bibinfo{volume}{85}},
  \bibinfo{pages}{4128} (\bibinfo{year}{2000}).

\bibitem[{\citenamefont{Green and Chamon}(2002)}]{Green:2002}
\bibinfo{author}{\bibfnamefont{D.}~\bibnamefont{Green}} \bibnamefont{and}
  \bibinfo{author}{\bibfnamefont{C.}~\bibnamefont{Chamon}},
  \bibinfo{journal}{Phys. Rev. B} \textbf{\bibinfo{volume}{65}},
  \bibinfo{pages}{104431} (\bibinfo{year}{2002}).

\bibitem[{\citenamefont{Rokhsar and Kivelson}(1988)}]{Rokhsar:1988}
\bibinfo{author}{\bibfnamefont{D.~S.} \bibnamefont{Rokhsar}} \bibnamefont{and}
  \bibinfo{author}{\bibfnamefont{S.~A.} \bibnamefont{Kivelson}},
  \bibinfo{journal}{Phys. Rev. Lett.} \textbf{\bibinfo{volume}{61}},
  \bibinfo{pages}{2376} (\bibinfo{year}{1988}).

\bibitem[{\citenamefont{Ardonne et~al.}(2004)\citenamefont{Ardonne, Fendley,
  and Fradkin}}]{Ardonne:2004uq}
\bibinfo{author}{\bibfnamefont{E.}~\bibnamefont{Ardonne}},
  \bibinfo{author}{\bibfnamefont{P.}~\bibnamefont{Fendley}}, \bibnamefont{and}
  \bibinfo{author}{\bibfnamefont{E.}~\bibnamefont{Fradkin}},
  \bibinfo{journal}{Annals of Physics} \textbf{\bibinfo{volume}{310}},
  \bibinfo{pages}{493} (\bibinfo{year}{2004}).

\bibitem[{\citenamefont{Henley}(2004)}]{Henley:2004ys}
\bibinfo{author}{\bibfnamefont{C.~L.} \bibnamefont{Henley}},
  \bibinfo{journal}{Journal of Physics: Condensed Matter}
  \textbf{\bibinfo{volume}{16}} (\bibinfo{year}{2004}).

\bibitem[{\citenamefont{Papanikolaou et~al.}(2007)\citenamefont{Papanikolaou,
  Luijten, and Fradkin}}]{Papanikolaou:2007kx}
\bibinfo{author}{\bibfnamefont{S.}~\bibnamefont{Papanikolaou}},
  \bibinfo{author}{\bibfnamefont{E.}~\bibnamefont{Luijten}}, \bibnamefont{and}
  \bibinfo{author}{\bibfnamefont{E.}~\bibnamefont{Fradkin}},
  \bibinfo{journal}{Physical Review B} \textbf{\bibinfo{volume}{76}}
  (\bibinfo{year}{2007}).

\bibitem[{\citenamefont{Castelnovo et~al.}(2005)\citenamefont{Castelnovo,
  Chamon, Mudry, and Pujol}}]{Castelnovo:2005cr}
\bibinfo{author}{\bibfnamefont{C.}~\bibnamefont{Castelnovo}},
  \bibinfo{author}{\bibfnamefont{C.}~\bibnamefont{Chamon}},
  \bibinfo{author}{\bibfnamefont{C.}~\bibnamefont{Mudry}}, \bibnamefont{and}
  \bibinfo{author}{\bibfnamefont{P.}~\bibnamefont{Pujol}},
  \bibinfo{journal}{Annals of Physics} \textbf{\bibinfo{volume}{318}},
  \bibinfo{pages}{316} (\bibinfo{year}{2005}).

\bibitem[{\citenamefont{Lee}(2007)}]{Lee:2007oq}
\bibinfo{author}{\bibfnamefont{S.-S.} \bibnamefont{Lee}},
  \bibinfo{journal}{Physical Review B} \textbf{\bibinfo{volume}{76}},
  \bibinfo{pages}{075103} (\bibinfo{year}{2007}).

\end{thebibliography}

\end{document}